\documentclass[aps,prl,amsmath,amssymb,amsfonts,showpacs,superscriptaddress,floatfix,twocolumn]{revtex4}
\usepackage{graphicx}
\newcommand{\be}{\begin{equation}}
\newcommand{\ee}{\end{equation}}

\newcommand{\bea}{\begin{eqnarray}}
\newcommand{\eea}{\end{eqnarray}}

\newcommand{\la}{\langle}
\newcommand{\ra}{\rangle}

\newcommand{\lp}{\left(}
\newcommand{\rp}{\right)}
\renewcommand{\vec}[1]{{\bf #1}}
\renewcommand{\epsilon}{\varepsilon}

\newcommand{\op}{\rho_{\pi}}
\begin{document}

\title{Synchronization and Dephasing of Many-Body States in Optical Lattices}
%%Synchronization and Dephasing in a Quantum Chain Far From Equilibrium}
\author{M.~B.~Hastings}
\affiliation{Center for Nonlinear Studies and Theoretical Division,
Los Alamos National Laboratory, Los Alamos, NM, 87545}
\affiliation{Kavli Institute for Theoretical Physics, University of California, Santa Barbara, CA 93106}
\author{L.~S.~Levitov}
\affiliation{Department of Physics, Massachusetts Institute of Technology,
77 Massachusetts Ave., Cambridge, MA, 02139}
\affiliation{Kavli Institute for Theoretical Physics, University of California, Santa Barbara, CA 93106}
\begin{abstract}
We introduce an approach to describe quantum-coherent evolution of a system of cold atoms in an optical lattice triggered by a change in superlattice potential. Using a time-dependent mean field description, we map the problem to a strong coupling limit of previously studied time-dependent BCS model.
We compare the mean field dynamics to a simulation using light-cone methods and find reasonable
agreement for numerically accessible times.
The mean field model is integrable, and gives rise to a rich behavior,
in particular 
to beats and recurrences in the order
parameter, as well as singularities in the momentum distribution, directly measureable in cold atom experiment.
\end{abstract}
\maketitle

The approach to thermal equilibrium has been a central problem in statistical mechanics since Boltzmann's H-theorem. 
Recent advances in ultracold atoms make it possible to probe
this approach in quantum-coherent many-body systems, due to the
ability to change interactions in optical lattices\cite{bloch1,bloch2} on a fast timescale.

These new experimental opportunities stimulated theoretical work on quantum dynamics in many-body systems.
Improvements in simulation
algorithms, such as time-dependent density-matrix renormalization group\,\cite{tebd,tdmrg}
and light-cone methods\,\cite{lc}, make it possible to study these
systems numerically for short times.
However, the increase in entanglement entropy limits the simulation time\,\cite{calab}, calling for the development of new approaches.

One simple-to-realize way to start a system out of equilibrium is to begin
with an additional period-two modulation in a translationally invariant
system.  One case of this is an XXZ spin chain started from a Neel state at large Ising
coupling \cite{calab}.  Another case
proposed recently is a Bose gas in an optical
lattice, with a period-two superlattice initially superimposed\cite{proposed},
causing the system to begin in a state with alternating filled and empty sites.
The superlattice is then removed, and the
system evolves under Bose-Hubbard dynamics.

In this article we study interacting spinless fermions in a one-dimensional lattice, described by the Hamiltonian 
\be
\label{fermi}
H=
%%LL \frac{t}{2} 
\Delta \sum_{i=1...N} 
\lp a_i^\dagger a_{i+1} + {\rm h.c.}\rp +
\sum_{i=1...N} 
\lambda \hat n_i \hat n_{i+1},
\ee
where  $\Delta$ is the hopping amplitude, $\hat n_i=a_i^{\dagger} a_i -\frac12$.
The initial state, taken to be alternating filled and empty sites, $\hat n_i=\pm\frac12$, is created by an additional period-two
potential that is removed at $t=0$, after which the system evolves
under $H$.
The Hamiltonian (\ref{fermi}) describes the regime in which multiple occupancy of lattice sites is inhibited by repulsive interaction and/or the Pauli principle.

To understand the evolution governed by (\ref{fermi}) we employ
a mean-field description of the problem (\ref{fermi}) which uses the staggered density 
\be\label{rho_pi}
\rho_\pi(t) = \frac{1}{N} \sum_{i=1...N} (-1)^i \langle \hat n_i \rangle
\ee
as an order parameter. 
We show that the resulting mean-field dynamics is mathematically equivalent to time-dependent BCS dynamics \cite{bcs1,bcs2,bcs3,bcs4},
with, however, very different initial conditions. Comparison to the numerical results\,\cite{lc} for the Jordan-Wigner-equivalent XXZ spin chain is used to test validity of the mean-field approach.

\begin{figure}
\centerline{
\includegraphics[scale=0.3,angle=270]{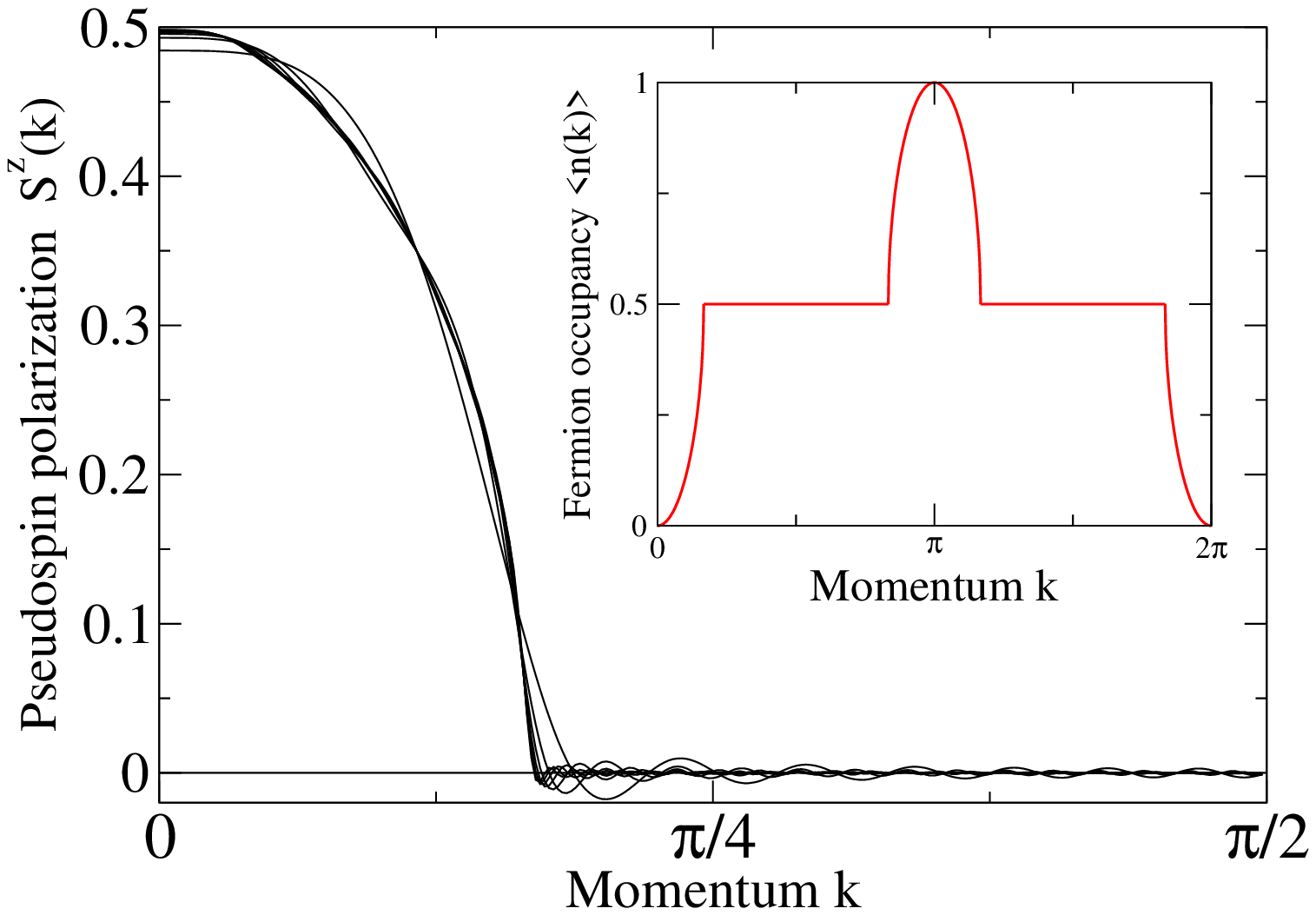}}
\caption{Buildup of singular momentum distribution of pseudospin $S^z_k$,  a semicircle (\ref{limitd}) with the edge at $k_c=\arcsin\lambda$.
A series of traces computed for $\lambda=0.5$ are shown at times $t=0,25,50,...,175$, eventually
converging to Eq.~(\ref{limitd}).
Inset: the corresponding limiting fermion momentum distribution $\la n(k)\ra =\la a^\dagger_k a_k\ra$ \cite{n(k)}. 
%% {\bf remove $<...>$ from the inset y label} 
}
\label{sz.5} 
%% \vspace{5mm}
\end{figure}

We find that instead of simple relaxation to a steady state, the order parameter time evolution exhibits revivals. These revivals are understood as resulting from a buildup of singularities in the fermion momentum distribution, illustrated in Fig.\ref{sz.5}. The formation of a steady state with a singular momentum distribution can be directly tested in a free flight imaging experiment \cite{Greiner2002}. 
%% by imaging particle density after free expansion \cite{Greiner2002}. 

The mean-field approximation can be constructed by replacing
$\hat n_i \hat n_{i+1} 
\approx
-2 \rho_\pi(t) \sum_i (-1)^i \hat n_i$ in the fermionic Hamiltonian (\ref{fermi}), which gives
\be
\label{mfH}
H_{\rm mf}=\frac{1}{2} \sum_i \lp a_i^\dagger a_{i+1} + {\rm h.c.}\rp
-2 \lambda \rho_\pi(t) \sum_i (-1)^i \hat n_i
\ee
(without loss of generality we set the hopping amplitude to $\Delta=\frac12$).
In the evolution governed by $H_{\rm mf}$ the quantity $\rho_\pi(t)$ is determined self-consistently via Eq.(\ref{rho_pi}).

To establish equivalence of the problem (\ref{mfH}) to the BCS dynamics \cite{bcs1,bcs2,bcs3,bcs4}
we consider a translation invariant
system on a ring, with momentum states $|k\rangle$, $-\pi<k<\pi$.  
Initially, the system is half-filled and each pair of momentum states, $|k\rangle$ and
$|k+\pi\rangle$, contains exactly one particle.
Momentum-nonconserving terms in the Hamiltonian
(\ref{mfH}) 
%% contains  that 
couple momenta $k$ and
$k\pm\pi$.
Thus, at all times
we can write the many-body wavefunction in a BCS-like form
\be
\Psi(t)=\prod_{-\pi/2 \leq k<\pi/2} \Bigl( u_k(t) a^{\dagger}(k+\pi)+v_k(t) a^{\dagger}(k)
\Bigr)|0\rangle,
\ee
where $|0\rangle$ is the vacuum state.
The initial conditions are
$u_k(0)=v_k(0)=\frac{1}{\sqrt{2}}$.
The evolution equations are
\begin{eqnarray}
\label{uvdyn}
\partial_t u_k(t)&=&i \cos(k) u_k(t)+2i\lambda \rho_\pi(t) v_k(t),
\\ \nonumber
\partial_t v_k(t)&=&-i \cos(k) v_k(t)+2i\lambda \rho_\pi(t) u_k(t),
\end{eqnarray}
with the self-consistency condition (\ref{rho_pi})
taking the form
\be
\label{scuv}
\rho_\pi(t)=\frac{1}{N/2} \sum_{-\pi/2\leq k <\pi/2}
{\rm Re}(\overline u_k(t) v_k(t)).
\ee
Note that there are $N/2$ different values of $k$ in the sum
in Eq.~(\ref{scuv}), so that $-1/2\le \rho_\pi(t)\le 1/2$.

Given a pair, $u_k,v_k$, we define pseudo-spins\,\cite{Anderson58} by
%% \be\label{Sxyz}
$
S^z_k ={\textstyle \frac12} (|u_k|^2-|v_k|^2),\quad
S^x_k+iS^y_k=\overline u_k v_k 
$
.
%% \ee
In terms of these classical variables the Hamiltonian reads
\be
\label{psH}
H_S=-\sum_{-\pi/2 \leq k<\pi/2}
2 \cos(k) S^z_k  +
\frac{2 \lambda}{N/2}\sum_{k,k'} S^x_k  S^x_{k'}
\ee
which, together with the usual Poisson brackets, $\{S_a,S_b\}=i\epsilon_{abc}S_c$,
reproduces the dynamics (\ref{uvdyn}).

The canonical BCS Hamiltonian \cite{Anderson58} differs from (\ref{psH}) in one important respect:
the BCS problem has an additional coupling $S^y_k  S^y_{k'}$.
We circumvent this problem in two steps.  First, we extend the Hamiltonian (\ref{psH}) to a twice larger momentum range $-\pi<k<\pi$, doubling the number of
$k$-states.
Simultaneously we replace the coupling $2\lambda/(N/2)$ by
$2\lambda/N$.  
Consider any pair of momenta, $k$ and $k+\pi$; these two pseudospins
feel opposite $z$ fields and the same $x$ field,
so they evolve as 
\[
S^x_{k+\pi}(t)=S^x_{k}(t),\quad
S^{y,z}_{k+\pi}(t)=-S^{y,z}_{k}(t),
\]
where individual spin polarizations for $-\pi/2<k<\pi/2$ evolve in the same way as
in the original problem (\ref{psH}).  

After the states are doubled, the net $y$ polarization $\sum_k S^y_k $ taken over the extended range $-\pi<k<\pi$ vanishes at all times, while the net $x$ polarization 
$\sum_k S^x_k $ does not change, and so we can then add
the interaction $S^y_k  S^y_{k'}$ to the original
Hamiltonian  (\ref{psH}) without changing the dynamics.  Thus, we arrive at
the BCS Hamiltonian:
\be\label{H_doubled}
H_{\rm BCS}=-\sum_{-\pi \leq k<\pi}
2 \cos k\, S^z_k  +
 \frac{2 \lambda}{N}\sum_{k,k'} S^x_k  S^x_{k'} +
S^y_k  S^y_{k'}.
\ee
With the initial state $S^x_k=1$, $S^{y,z}_k=0$, the problem
(\ref{H_doubled}) yields the dynamics of $\rho_\pi(t)$ identical to (\ref{psH}).

\begin{figure}
\centerline{
\includegraphics[scale=0.3,angle=270]{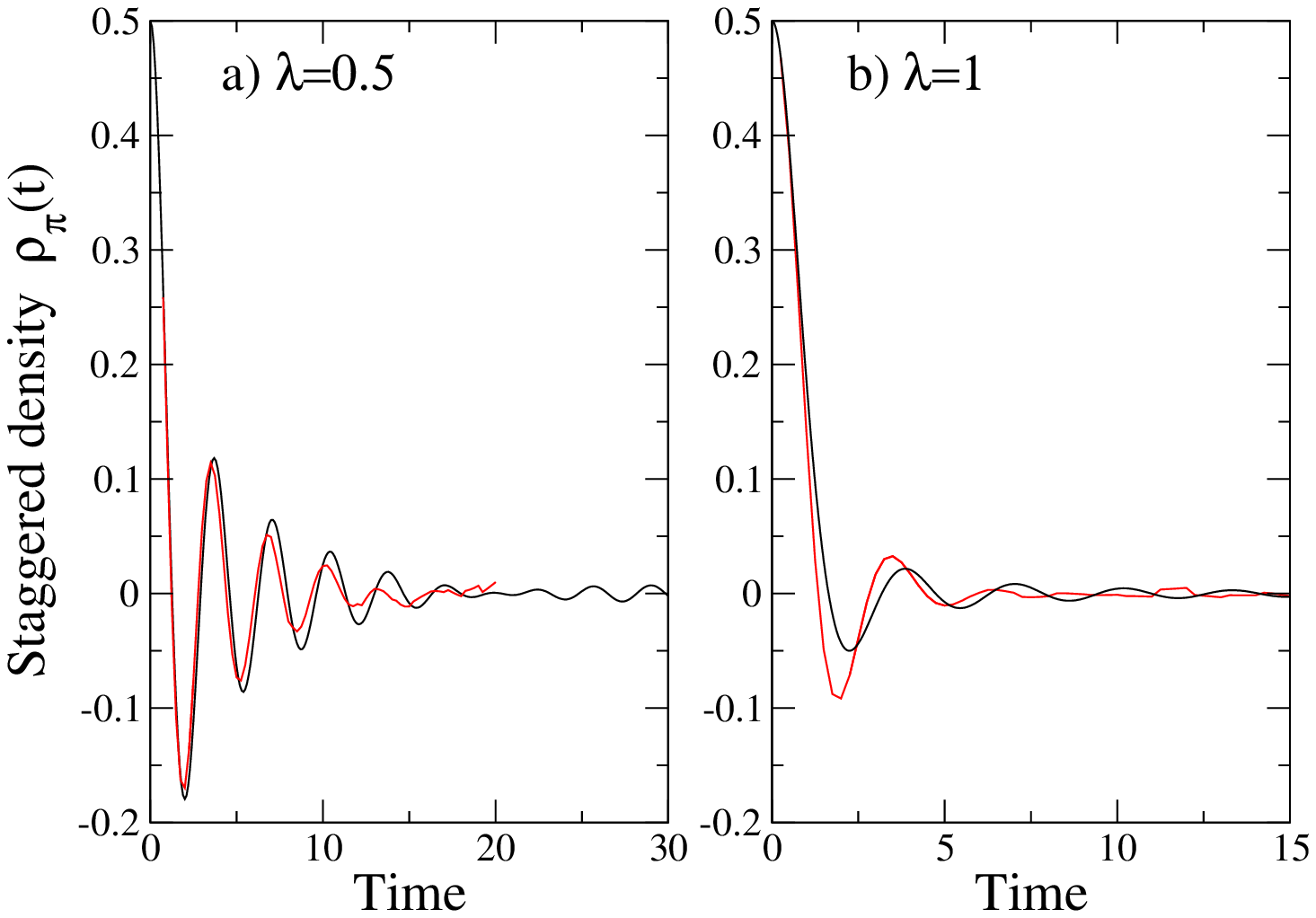}}
\caption{Comparison of mean-field (black, Eq.\eqref{rho_pi}) and exact dynamics (red, Eq.\eqref{M(t)}) simulated
using light-cone methods\,\cite{lc}.
The revival in a) at $t\gtrsim 20$ can be modeled by Eq.\eqref{beating}.
%% ; (a), (b) is $\lambda=0.5,\,1$ respectively.
%% {\bf increase lettering (a) $\lambda=0.5$ and (b) $\lambda=1$}
}
\label{comp} 
%% \vspace{5mm}
\end{figure}

As a sanity check of the mean-field approximation we use comparison to the simulation of the XXZ spin-$\frac12$ chain with the $zz$ coupling of strength $\lambda$: $H_{\rm XXZ}=\sum_i S^x_iS^x_{i+1}+ S^y_iS^y_{i+1}+\lambda S^z_iS^z_{i+1}$
which is Jordan-Wigner-equivalent to (\ref{fermi}).
The initial conditions are taken to be the Neel state: alternating spin
up and down.  The staggered density (\ref{rho_pi}) translates to the Neel order parameter 
\be\label{M(t)}
M(t) = \frac{1}{N} \sum_i (-1)^i \langle S^z_i(t) \rangle
.
\ee
In Fig.~\ref{comp}(a) we compare the XXZ simulation
done using the light-cone method \cite{lc}
to the mean-field
dynamics (\ref{uvdyn}) for $\lambda=0.5$.  The mean-field dynamics was simulated using 4th order Runge-Kutta
with 25000 pairs of modes $u_k$, $v_k$ with $0\le k<\pi/2$
and a time step of $0.025$; the parameters were chosen to assure
insensitivity of the
results to finite size and finite timestep. The behavior of the exact $M(t)$ and the mean-field $\rho_\pi (t)$ is quite similar, with $M(t)$ oscillating slightly faster than $\rho_\pi (t)$.
The later times of the light-cone dynamics show noise from sampling errors.
Note that at times of just past 20, the mean-field dynamics shows a revival:
the amplitude of oscillations begins to increase again.  
This revival
is not far outside the times reached with the light-cone methods, and may
be accessible with more numerical effort.

As we will see below, for $\lambda>1$ the mean-field dynamics 
predicts a non-vanishing asymptotic value $\op(t\to\infty)$,
while simulation of the XXZ chain indicates that $\op(t)$ rapidly decays to zero.
%% \mpar{This is strange... Still strange...}
Nevertheless, as illustrated in Fig.~\ref{comp}(b), even at $\lambda=1$ there is a reasonably good agreement between the two approaches at short times,
%% the mean-field dynamics agrees reasonably well with the light-cone results, 
with both 
$M(t)$ and $\op(t)$ decaying faster than for $\lambda=0.5$.

Turning to discuss different regimes, we note that because both 
the Hamiltonian (\ref{mfH}) and the initial values
$u_k(0)$ and $v_k(0)$ are real, the mean-field dynamics of $\op(t)$ is time reversal invariant.
Further, the initial state is invariant under the
orthogonal operator $O=\prod_k S^x_k$, which anti-commutes with the first term in $H_{\rm mf}$ and
commutes with the second.  Combining
these two statements, we see that the
dynamics depends only on the magnitude of $\lambda$ and not on its sign
(all this is also true for the XXZ chain started in the Neel state, where we set $O=\prod_{i \,{\rm odd}}S^z_i$). 

In the absence of interaction, $\lambda=0$, the decay of $\op(t)$ follows 
$\cos(2t + \delta)/t^{1/2}$ with $\delta=\frac{\pi}4$\cite{proposed,lc}. 
At $0<|\lambda|<1$, the system
is in a ``dephasing" regime, and
$\op$
is well-described at long times by
a sum of two frequencies beating together with a power-law
envelope decaying as $t^{-3/2}$:
\be\label{beating}
\op(t) \sim \Bigl(
a_1 \cos(\omega_1 t + \delta_1) +
a_2 \cos(\omega_2 t + \delta_2)\Bigr)/t^{3/2}.
\ee
We will see that this unexpected behavior, leading to revivals in $\op(t)$, signals formation of a singularity in $k$ space of the asymptotic polarization $S^z_k $. Below, we use integrals of motion of the BCS dynamics to show, in agreement with numerics, that $\omega_1=2$, $\omega_2=2\sqrt{1-\lambda^2}$.

One can expect that strong interaction will stabilize the state with density modulation (\ref{rho_pi}). In agreement with this intuition, we find that
for $|\lambda|>1$ the system is in the ``polarized" regime,
with non-vanishing $\op(t\to\infty)=(1/2)\sqrt{1-1/\lambda^2}$. 
We obtain
this asymptotic value analytically,
and confirm it numerically. The approach to the asymptotic value
is described by
$
\op(t)-\sqrt{1-1/\lambda^2} \sim \cos(\omega t + \delta)/t^{1/2}
$.
At the phase transition at $\lambda=1$, we observe numerically that
$
\op(t) \sim \cos(2 t+\delta)/t^{3/2}
$. 

The behavior in the dephasing regime can be qualitatively understood as follows.
In the non-interacting case, all of the pseudo-spins remain in the $x-y$
plane, with $S^z_k(t)=0$.  Initially, they all point in the
$x$-direction, and dephase over time, leading to the $1/\sqrt{t}$ decay
in $\op$.  When interaction is turned on, the spins begin to move
out of the $x-y$ plane and polarize in the $z$-direction.  The appearance of
a net $z$-polarization is required by conservation of energy: as the ferromagnetic $S^x S^x$ term in the energy (\ref{mfH}) is
becoming less negative because of dephasing, the $S^z$ contribution must become
more negative.

However, the asymptotic distribution of $S^z$ is {\it not} the thermal
distribution.  Instead, it shows a square-root singularity at
$k=k_c \equiv\arcsin\lambda$ (see Fig.~\ref{sz.5}).
We find, analytically and numerically (see Fig.\ref{sz.5}),
that $S^z_k  (t\rightarrow\infty) = 0$ for $|k|>k_c$, whereas
\be
\label{limitd}
S^z_k(t\rightarrow\infty)= {\textstyle \frac12} \sqrt{\cos^2 k\,-1+\lambda^2},
\quad |k|\leq k_c
.
\ee
The van Hove singularity at the band edge and the 
singularity at $k_c$ give rise to the frequencies $\omega_1$, $\omega_2$
in (\ref{beating}).

An analytic insight into the behavior in the dephasing regime can be gained as follows.
The BCS dynamics has infinitely many commuting integrals of motion that can be written as an energy dependent 
Lax vector \cite{bcs2}
%
%% \be\label{L(xi)}
\[
\vec L(\xi) = \hat{\vec z}+2 \lambda{\sum}_{\xi'\ne \xi}\,\frac{\vec S_{\xi'}}{\xi-\xi'}
,
\]
%% \ee
%
where in our case $\xi=\cos k$ and $\sum_{\xi'} ...\vec S_{\xi'} =\int ...\vec S_{k}\frac{dk}{2\pi}$.

The asymptotic polarization can be found by comparing the values $\vec L(\xi)$ in the initial and asymptotic states \cite{bcs4}.
The initial state, polarized in the $x$ direction, gives
\be\label{Linitial}
\vec L^2(\xi) = L_1^2+L_3^2=1+\lambda^2/(\xi^2-1)
.
\ee
In the asymptotic state, because the $x$ and $y$ components are dephased, we can approximate:
\be\label{Lfinal}
\vec L^2(\xi) \approx L_3^2 = \lp 1+2\lambda\int_{-\pi}^\pi \frac{S^z_k  dk}{(\xi-\cos k)2\pi}\rp^2
.
\ee
Comparing (\ref{Linitial}) and (\ref{Lfinal}) we obtain an integral equation 
\be\label{IntegralEquation}
2\lambda\int_{-\pi}^\pi \frac{S^z_k  dk}{(\xi-\cos k)2\pi} =
\sqrt{\frac{\xi^2-1+\lambda^2}{\xi^2-1}} - 1
\ee
where $\xi$ is treated as a complex variable, ${\rm Im}\,\xi>0$.
Changing variable to $x=\cos k$, we rewrite (\ref{IntegralEquation}) as
\be\label{eq:Int_Eqn}
\frac1{2\pi}\int_{-1}^{+1}\frac{f(x)dx}{\xi-x} = \sqrt{\frac{\xi^2-1+\lambda^2}{\xi^2-1}} - 1
\ee
where $f(x)=4\lambda S^z(x)/\sqrt{1-x^2}$ is the unknown function.
This equation can be solved using Cauchy's
formula, by writing
$f(x)=f_+(x)+f_-(x)$, the  functions $f_\pm(x)$ being analytic in the
upper/lower complex halfplane, respectively.  The result is
$f(x)= 2\,{\rm Im}\,\sqrt{\frac{x^2-1+\lambda^2}{x^2-1}}$, which
yields the semicircle dependence (\ref{limitd}).

\begin{figure}
\centerline{
\includegraphics[scale=0.3,angle=270]{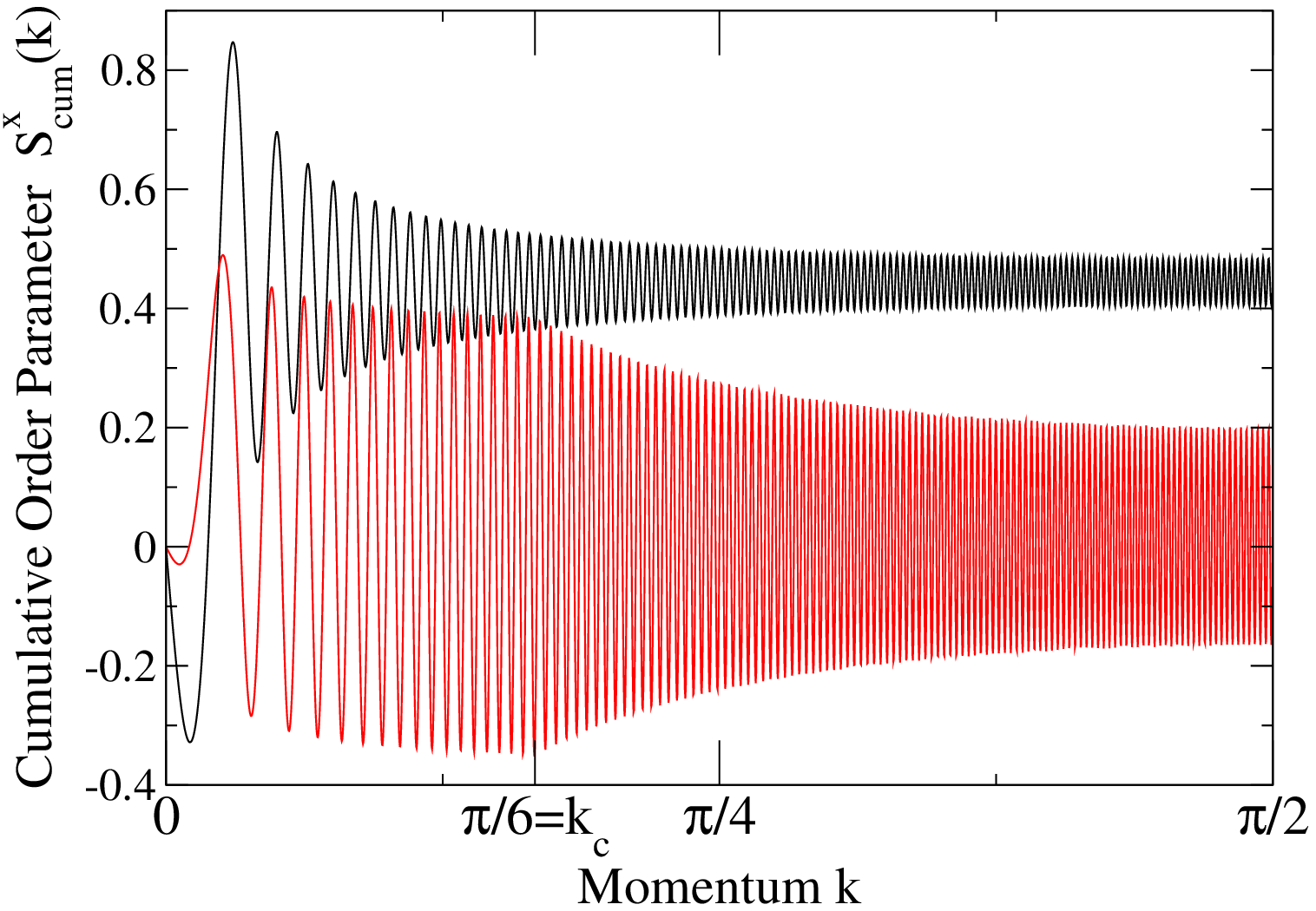}}
\caption{
%% Cumulative order parameter 
The quantity $S^x_{\rm cum}(k)=\frac1{N} \sum_{0<k'<k} S^x_{k'}$, 
%% Eq.(\ref{Scumul}), 
for $\lambda=0$ (black) and $\lambda=0.5$ (red). Snapshots at $t=400$
are shown, with $y$-axis 
multiplied by an arbitrary scaling factor in both cases.}
\label{scan} 
%% \vspace{5mm}
\end{figure}

The $t^{-3/2}$ power law envelope in (\ref{beating}) 
is more difficult to understand.  Consider the band edge singularity at $k=0$.
At $\lambda=0$,
this gives rise to a $1/\sqrt{t}$ contribution to $\op$ due to a coherent
contribution of the modes with $k\lesssim 1/\sqrt{t}$.  
For $\lambda>0$,
the $x-y$ component of the spins near $k=0$ is proportional to $k$, which
suggests in a scaling picture 
that $S^x(0)$ should decay with an envelope
of $1/\sqrt{t}$, which we do observe numerically.
Thus, if the order parameter were due to
a coherent contribution of spins with $k \lesssim 1/\sqrt{t}$,
we would expect a $1/t$ decay in $\op$, rather than $1/t^{3/2}$ as observed.
To understand this better, we analyze
the cumulative order parameter $S^x_{\rm cum}(k)$, plotting it 
% \be\label{Scumul}
% S^x_{\rm cum}(k)=(1/N) \sum_{0<k'<k} S^x_{k'}
% .
% \ee  
% We plot $S^x_{\rm cum}$ {\it vs.} $k$ 
in Fig.~\ref{scan} for
a particular snapshot in time.
It can be seen that the contribution of the low-lying modes is
largely canceled out by higher modes in the interacting
case, while the contribution of the higher modes averages
out in the non-interacting case.

To identify the boundary of the dephasing regime, we recall that in the BCS problem
the large-$t$ asymptotic is governed by the complex roots of the spectral equation $\vec L^2(\xi)=0$ \cite{bcs2,bcs3,bcs4}. 
%% The number of these roots is in a one-to-one relation with the asymptotic behavior of $\Delta(t)$. 
In our case, for the initial state polarized in the $x$-direction, 
the spectral equation can be factorized as
$\vec L^2=(L_3+iL_1)(L_3-iL_1)=0$, giving
\be
1\pm i 2 \lambda\int_{-\pi}^\pi \frac{dk}{4\pi(\xi-\cos k)} = 1\pm  \frac{\lambda}{\sqrt{1-\xi^2}} = 0
\ee
This equation has complex imaginary roots $\xi=\pm i \sqrt{1-\lambda^{-2}}$ if $\lambda>1$, and has no complex solutions for $\lambda<1$. 
Thus, for $\lambda>1$, we have a non-zero asymptotic value of
$\op=\frac12 \sqrt{1-\lambda^{-2}}$, which we confirmed by
numerical simulations.

We have tried non-integrable deformations of the model, by making the
$S^x S^x$ coupling between modes weakly dependent on $k$.
Even a very small change removes the singularity at $k_c$, though
a smooth kink remains, causing the contribution
to $\op(t)$ at frequency $\omega_2=2\cos k_c$ to decay exponentially in time, rather
than as a power law.  At early times, beats are still observable.

Now we briefly discuss application of our results to bosons in an optical lattice, described by \cite{Jaksch98}
\be
\label{BoseHubbard}
H_{\rm BH}=
\Delta \sum_i \lp b_i^\dagger b_{i+1} + {\rm h.c.} \rp +
\sum_i U (b^{\dagger}_i b_i)^2
.
\ee
As above,
the initial state of alternating filled
and empty sites is imposed by an additional period-two potential.

Because the mean-field dynamics with the staggered
density order parameter $\rho_{\pi}(t)=
\frac{1}{N} \sum_i (-1)^i \langle b^{\dagger}_i b_i - 1/2 \rangle$
does not depend on statistics, the bosonic and fermionic Hamiltonians (\ref{BoseHubbard}), (\ref{fermi}) yield the same mean-field evolution.
In fact, in the bosonic case, 
for any initial conditions where we alternate a site with
$n$ particles with an empty site, we obtain the same mean-field equations
up to rescaling 
$U$ by dividing it by $n$.  We expect the mean-field
to become more accurate for larger $n$.

As a result, the mean-field theory for the bosonic system predicts
the asymptotic momentum distribution
$n(k)=\langle b^\dagger_k  b_k \rangle$ which is similar to fermionic
$n(k)$ (see Fig.\,\ref{sz.5}).
This momentum distribution can be measured in
a cold atom experiment using time-of-flight.

Unlike the fermion problem (\ref{fermi}), the Bose-Hubbard model (\ref{BoseHubbard}) is
not integrable, and therefore at long times the momentum distribution
of the particles should thermalize. Still, at short times the kink in the
distribution $n(k)$ may be observable. If present, it will manifest itself also in revivals of the staggered density amplitude $\op(t)$.

As the interaction $U$ increases,
it is no longer valid to use
mean-field theory.  However, because at $U=\infty$
the bosonic problem
reduces to non-interacting fermions, for large $U$ we can use
second order perturbation theory to map the problem onto a system
of hard-core bosons with weak attractive interactions
$\frac{4 \Delta^2}{U} \sum_i b^{\dagger}_i b_i b^{\dagger}_{i+1} b_{i+1}
+\frac{2 \Delta^2}{U} \sum_i b^{\dagger}_i b_i b^{\dagger}_{i+1} b_{i-1} + {\rm h.c.}$
For large $U$, both
terms are now weak and can be treated by mean-field theory.  The treatment
of the first term is as before, whereas the second term leads to a momentum dependent coupling in the BCS
mean-field which breaks integrability.

In contrast to that, the fermion problem is integrable, and thus its dynamics is not be ergodic. Thus our main results, the revivals in the order parameter and the formation of a singular momentum distribution, may persist in the fermion case even at the times longer than those described by our mean-field approach.

We thank J. Eisert for useful discussions.
MBH was
supported by U. S. DOE Contract No. DE-AC52-06NA25396.
LL's work was partially supported by W. M. Keck
Foundation Center for Extreme Quantum Information
Theory and by the NSF grant PHY05-51164.


\begin{thebibliography}{99}
\bibitem{n(k)}
The $z$-component $S^z_k $ is related to the density of particles with momentum
$k$ by $\la n(k)\ra
=\frac12-S^z_k $ for $-\pi/2\leq k
<\pi/2$ and by $\la n(k)\ra =
\frac12+S^z_{k+\pi}$ otherwise.
\end{thebibliography}
\end{document}